\documentclass[12pt]{iopart}
\usepackage{iopams}
\usepackage{graphicx,graphics} % Allow PDF or EPS
\usepackage{supertabular}
\usepackage{times}
\usepackage{bm} % bold Greek
\usepackage{subfigure} %imposes to much vertical space
\usepackage{ulem}
\begin{document}

\title[Longitudinal  polariton condensation and superradiant emission]{Longitudinal  polariton condensation and superradiant emission in semiconductor edge-emitting laser structures}

\author{D L Boiko,$^{1}$ P P Vasil'ev,$^2$ }

\address{$^1$Centre Suisse d'Electronique et de Microtechnique SA, 2002,
Neuch\^atel, Switzerland \\
$^2$P N. Lebedev Physical Institute, 53 Leninsky prospect, Moscow 119991, Russia }
\ead{dmitri.boiko@csem.ch}
\begin{abstract}
We show that superradiant emission (SR) in semiconductor laser diode structures is governed by the master equation of the form of Ginzburg-Landau equation or Gross-Pitaevskii equation for a phase transition to coherent matter state. We conjectured condensation of one-dimensional longitudinal lower polaritons with effective mass of 10$^{-11}$ the mass of free electron.  Two different regimes of SR emission and polariton condensation are predicted, one of which exhibits some characteristic features of the Dicke superradiance in atomic gases. The predictions of analytic model  are confirmed by the results of numerical simulations based on Maxwell-Bloch equations, utilizing InGaN/GaN heterostructure quantum well as a model system.
\end{abstract}

%\submitto{\NJP}
\maketitle

\section{Introduction}

The search for spontaneous build-up of macroscopic coherences in semiconductor microcavities has resulted in intensive research towards a polariton laser. Interestingly, group III - nitride semiconductors
are capable of reaching macroscopic coherences at room temperature conditions. Thus due to large reduced exciton mass, the InGaN/GaN quantum well (QW) exciton binding energy is higher as compared to GaAs/AlGaAs QWs or other conventional III-V counterparts. Furthermore, due to a stronger exciton-photon Rabi coupling, InGaN/GaN microcavities offer possibility of polariton lasing at room temperature conditions \cite{Christopoulos07}. High critical temperature of two dimensional (2D) BEC transition is conditioned by
low in-plane effective mass of
polaritons, so as the critical density is reached before destruction of excitons. Such 2D BEC is a transient dynamic state, which can nevertheless be analyzed using a stationary Schr\"{o}dinger equation \cite{Boiko08}.

The superradiance (SR) in semiconductor edge emitters can be considered as %is
another example of spontaneous macroscopic coherences in solids \cite{Vasil'ev09}.
The cooperative radiative recombination in an ensemble of quantum oscillators (e.g. atoms or molecules)
has been predicted before the invention of lasers \cite{Dicke54}. Since that, it has been extensively studied both theoretically and experimentally \cite{Skribanowitz73,Schuurmans81}. The characteristic features of the SR emission are the temporal and spatial coherence, highly anisotropic emission pattern, quadratic dependence of pulse intensity on the number of excited atoms $I \propto n^2 $,
afterpulse ringing attributed to Rabi-type oscillations. The SR pulse duration decreases with the§ number of emitters $\tau_{sr} \propto 1/ n$, while the pulse energy is proportional to the ensemble population ($I\tau_{sr}\propto n $), in agreement with the energy  conservation considerations. On the other side, the spontaneous nature of the transition to the transient macroscopically coherent state is responsible for large fluctuations in the shape, duration and amplitude of SR pulses due to quantum-mechanical uncertainties  \cite{Andreev93,Gross82,Men'shikov99}.

In semiconductors, a hypothesis has been drawn that the SR is assisted by formation of a transient BCS-like state of electron-hole (e-h) pairs mediated by photons
%, much like in a
%superconducting BCS (Bardeen, Cooper, and Schrieffer) state of electron pairs   mediated by phonons
\cite{Vasil'ev01,Vasil'ev04}. %[9,10]. T
According to that,  the e-h system undergoes a second order non-equilibrium phase transition when the coherent e-h BSC-like state is building up during SR pulse emission.

In this paper, starting from  semiclassical travelling wave Maxwell-Bloch equations, we show that in semiconductors, superradiance regime is governed by the Ginzburg-Landau equation (GLE) or Gross-Pitaevskii master equation for a phase transition to coherent matter state. The effective mass of condensing one-dimensional (1D) longitudinal polaritons is very low, 10$^{-11}m_0$, giving a good reasoning for the thermal de Broglie wavelength criteria for macroscopic quantum degeneracy
%does not enter
 being not impacting our GLE for SR.  This  allows us to define the critical density in the system, the order parameter and the coherence length without accounting for the system temperature in the first approximation. These arguments thus further support the hypothesis of BCS-like condensation during SR emission in semiconductors. Our  GLE for SR in semiconductors has two types of solutions, indicating that two different regimes of SR emission and polariton condensation are possible. One of them (type-I SR) reveals features that are similar  to the Dicke superradiance in atomic gases with long natural decoherence time. In this regime, the macroscopic coherence is established over the entire sample. In the type-II SR, only partial coherence can be achieved in the sample so as the domain of condensate fraction is smaller than the size of the sample. SR pulse parameters predicted by analytic model are confirmed in numerical simulations based on the travelling wave  Maxwell-Bloch equations. Inspired by remarkable features of the group III - nitride microcavities in reaching room temperature BEC macroscopic coherences, we conducted our numerical studies of the SR in an edge-emitting ridge-waveguide cavity with InGaN/GaN QWs.

A serious study of cooperative coherent effects in atomic (molecular) vapor has been started %essentially
 in 1970s, 20 years after the predictions of Dicke \cite{Men'shikov99}. These studies revealed that SR emission,  generation in masers,  gyrotrons are of the same nature. The SR in small-size and extended samples has been studied.
The type-II SR regime in semiconductor edge emitting devices predicted here might be considered  by someone to be the Dicke SR regime in atomic gasses in case of an extended, optically thick medium \cite{MacGillivray76}. However the differences are (i) in the relationship between the decoherence time and SR pulse width and (ii) the presence of  a waveguide in semiconductor edge-emitting cavity. As a result,   the Dicke SR in extended atomic gasses reveals the pulse intensity  $\propto n^2$ \cite{Men'shikov99}, while our predictions for the pulse intensity in type-II SR regime in semiconductor cavities is $\propto n^{3/2}$, the result which we confirm by numeric simulations.

Because of low effective mass of longitudinal %lower
polaritons, we concluded that thermal potential would provide a negligible correction to the critical density for SR emission and polariton condensation. In 70s, there was quite large number of sophisticated calculations for SR phase transition in atomic or molecular gasses  that specifically look at the critical temperature of the transition \cite{Pimentel75}. Following Dicke, they treat an ensemble of atoms (or molecules) as a correlated spin states
%, in applications to a maser
\cite{Hepp73}. Unfortunately, most of these theoretical works are limited to obtaining a theoretical expression for the critical temperature $T_c$, without attempting to make
an estimate.
%Following along the lines of these studies, we shall replace our Eq.(\ref{Ncr}) by (see e.g.  Eq.(32) from Ref. \cite{Wang73})
Following along the lines of these studies, (e.g., examining   Eq.(32) from Ref. \cite{Wang73}),  we conclude that the thermal potential can be accounted for in our expression for the critical density  (\ref{Ncr}) as
\begin{equation}
\frac{\Gamma g_0 (n_{cr}^*-n_t)}{v_g(\alpha_i{+}2/L^*)} {=}\rm{tanh^{-1}(\frac 1 2 \frac{ \hbar \omega} {k_B T_c})},
\label{Ncr_Tcr}
\end{equation}
 The correction term from the thermal potential in the r.h.s of this expression $\propto 2 \exp(\frac{ - \hbar \omega} {k_B T_c})$  is negligible, about 10$^{-51}$.

The thermal potential $k_B T$ does not enters explicitly in threshold conditions for laser or maser equations, and this well agree with experimental observations. In Ref. \cite{Graham70}, a possibility to derive  GLE for a laser has been evoked. This conjecture is inline with our findings for the SR GLE (\ref{GLE}) that does not include thermal potential $k_B T$.

The paper is organized as follows. In Sec. \ref{num_mod}, we introduce traveling wave Maxwell-Bloch equations used as a starting point of our development and in the numeric simulations. In Sec. \ref{Analytic}, we obtain our GLE for SR and longitudinal polariton condensation in semiconductors. We analyze its two possible solutions. In  Sec. \ref{Results}, we provide a detailed discussion to our findings.

\section{Model system}
\label{num_mod}

The SR in semiconductors is usually achieved in an edge emitting laser diode cavity configuration \cite{Vasil'ev99}. Such cavity has a waveguide structure directing the emission of spontaneous photons and lasing modes.  Herewith we also consider the built up of SR in a semiconductor cavity pumped high above the transparency conditions of the semiconductor gain medium (bulk or quantum wells separate confinement heterostructure). Thus if during the  build-up of SR pulse, a phase transition to macroscopic quantum state takes place, the condensation is essentially one-dimensional.

In practice, to prevent the onset of lasing emission, % prior to the SR,
a tandem cavity configuration with an additional absorber section is  used \cite{Vasil'ev01}. The condensation takes place in the gain section, while  the absorber section is just an auxiliary arrangement for experimental observation. It does not impact the physics of SR emission %buildup
in strongly pumped semiconductor gain medium. Therefore, we limit development of our analytic theory to the processes in the amplifying semiconductor medium.

As a starting point, we use
a semi-classical Maxwell-Bloch  %approach for the
equations for the traveling wave amplitudes, carrier densities and macroscopic coherences (polarization) %which build up
in the ensemble of electrons and holes. (Similar equations can be found in studies of SR in atomic ensemble, in the case of extended source  \cite{MacGillivray76}.)
The evolution of the amplitudes of the forward ($A_+$) and backward ($A_-$) travelling waves %traveling in the Z-axis direction
is defined by the cavity loss and macroscopic medium polarization
\begin{equation}
%\begin{split}
\frac{\partial A_\pm}{\partial t } \pm v_g\frac{\partial A_\pm}{\partial z }= \frac 1 2 \Gamma \sqrt{\frac {g_0}{T_2} }P_\pm-\frac 1 2 v_g \alpha_i A_\pm,
%
%\frac{\partial A_\pm}{\partial t } \pm v_g\frac{\partial A_\pm}{\partial z }= \Gamma \frac{2 \pi \omega v_g^2}{c^2}P_\pm-\frac 1 2 v_g \alpha_i A_\pm,
%
%\end{split}
\label{TravAmpl}
\end{equation}
where the field amplitudes are normalized in the secondary quantization convention $\hat{a}_\pm\left|N_\pm\right\rangle{=}A_\pm\left|N_\pm{-}1\right\rangle$:
\begin{equation}
%\begin{split}
\textbf{E}_\pm{=}\sqrt{\frac{4 \pi \hbar \omega v_g^2}{c^2}} \textbf{e}_\pm A_\pm \sin(\omega t \mp kz),\quad A_\pm{=}\sqrt{N_\pm}
%\end{split}
\label{NormA}
\end{equation}
The same
normalization used for coherences (the ensemble average for the off-diagonal density matrix elements) induced between the electrons and holes and associated with the forward and backward waves reads
\begin{equation}
%\begin{split}
%initial
%\textbf{P}_\pm{=}\sqrt{\frac{c^2 g_0 }{4 \pi \hbar \omega v_g^2 T_2}} \textbf{e}_\pm P_\pm \cos(\omega t \mp kz)
%typo_corected \hbar
\textbf{P}_\pm{=}\sqrt{\frac{c^2 \hbar g_0 }{4 \pi \omega v_g^2 T_2}} \textbf{e}_\pm P_\pm \cos(\omega t \mp kz).
%\end{split}
\label{NormP}
\end{equation}
Here the $z$ axis is directed along the cavity waveguide.  As shown below, the macroscopic variables $P_\pm$ %would
provide a measure for the order parameter of the system.
The dynamics of carrier populations and coherences is described as follows:
\begin{eqnarray}
%\begin{equation}
%\hspace{-0.1in}\begin{split}
&\frac{\partial P_\pm}{\partial t }{=}{-}\frac{P_\pm}{T_2}
%{+}D\frac{\partial^2 P_\pm}{\partial z^2}
{+}\sqrt{\frac{g_0}{T_2}}(n{-}n_t) A_\pm
%\\
%&\quad\quad{+}\sigma\sqrt{\frac{g_0}{T_2}}(n_a{-}n_V) A_\pm
{+} \Lambda_\pm \\
&\frac{\partial n}{\partial t }{=}
%D\frac{\partial^2 n_\pm}{\partial z^2}
{-}\frac{n}{\tau_n}{-}\sqrt{\frac{g_0}{T_2}}(A_{+}P_{+}{+}A_{-}P_{-}){+}\frac{J(z{,}t)}{e_q d}
%\\
%&\frac{\partial n_a}{\partial t }{=}
%%D\frac{\partial^2 n_a}{\partial z^2}
%{-}\frac{n_a}{\tau_a}{-}\sigma\sqrt{\frac{g_0}{T_2}}(A_{+}P_{+}{+}A_{-}P_{-})
\label{P&n}
%\end{split}
%\end{equation}
\end{eqnarray}
where
the Langevin force term  due to polarization (thermal) noise $\Lambda_{\pm}$ triggers the spontaneous build-up of superradiant pulse. This term is important for numerical solutions, while semiclassical analytical theory we develop here does not require it.

  In Eqs.(\ref{TravAmpl})-(\ref{P&n}), $v_g=c/n_g$ is the group velocity of modes  in the cavity waveguide, $g_0$ is the differential gain coefficient,  $\tau_n$ and $T_2$ are the carrier relaxation time and decoherence time,  the coefficient $\Gamma$ accounts for the partial overlap of the waveguide mode field and the gain medium (e.g. quantum wells), $n_t$ is the carrier density at transparency. The coefficient  $\alpha_i$ accounts for the optical loss in semiconductor medium, $J(z,t)$ is the pump current density and $d$  is the thickness of the active region (e.g. $d=d_{QW} N_{QW} $ in a heterostructure with $N_{QW}$ QWs of thickness $d_{QW}$). We neglect the carrier diffusion since it is not significant at the time scale set by the width of the SR pulse ($\sim$ 1 ps).

 For spontaneous emission, the coherence (polarization of the medium) decays faster then the population inversion.
In that case, the medium polarization follows adiabatically the optical field ($\partial P_{\pm}/\partial t,\partial P_{\pm}/\partial z=0$).
The equation for polarization can be adiabatically excluded so as the model transforms into the standard travelling wave model for ultrafast semiconductor lasers.
 However in the case of SR, this adiabatic approximation is not valid.

 Thus, we use the same set of approximations
and  parameters as used in the traveling wave rate equation model of a semiconductor laser
 but without adiabatic approximation  for the macroscopic polarization. In particular, the details of the vertical composition of epitaxial layers do not enters the model directly.

 In what follows, the model calculations are performed assuming InGaN/GaN double quantum well separate confinement heterostructure with the usual arrangement of epitaxial layers for an edge emitting laser \cite{Dorsaz11,Kuramoto10}. The main parameters of the model are summarized in Table \ref{Param}.

\Table{\label{Param} Model parameters for wide-bandgap semiconductors, using data from Ref.\cite{Scheibenzuber11,Dorsaz11,Smetanin11}.}
%\begin {table}[t]
%\caption {Model parameters for SR emission in wide-bandgap semiconductors, using data from Ref.\cite{Scheibenzuber11,Dorsaz11, Smetanin11}.}
%\label {Param}%[th]
%%\begin {indented}
%%\item []
%\begin {ruledtabular}
%%\begin {tabular}{llllll}
%\begin {tabular}{lll}
%%& array size & 2x2 & 3x3 & 4x4 & 5x5 \\
\br
Parameter  & Name  & Value  \\
\mr
$d_{QW}$ & QW width & 3 nm
%\multicolumn {1}{c}{77.4} & \multicolumn {1}{c}{185} &
%\multicolumn {1}{c}{338.6 } & \multicolumn {1}{c}{538.24}
\\
%\hline
$N_QW$ & number of QWs & 2
%\multicolumn {4}{c}{0.58}
\\
$n_g$ & group refractive index $c/v_g$ & 3.5
%\multicolumn {4}{c}{ 0.40}
\\ %\hline
$w$ & waveguide width & 2 $\mu$m
%\multicolumn {4}{c}{1.611}
\\
$\Gamma$ & optical confinement factor
& 0.02
%\multicolumn {4}{c}{7.69\footnotemark [1] }
\\
$n_t$ & carrier density at
transparency
& 1.6$\cdot$10$^{19}$ cm$^{-3}$
%\multicolumn {4}{c}{2}
\\
%\hline
$\tau_n$ & carrier lifetime (gain section) &
0.9 ns
%\multicolumn {4}{c}{13}
\\
%$\tau_a$ & carrier lifetime (absorber)  &
%5 ps
%%\multicolumn {4}{c}{266.1}
%\\
<$T_2$ & carrier decoherence  time & 100 fs
%\multicolumn {4}{c}{92.1%
%\footnotemark [1] }
\\
$g_0{=}\partial g / \partial n$ & differential material gain &
1.2$\cdot$10$^{-6}$ cm$^{3}$/s
%\multicolumn {4}{c}{5184}
\\
%\end {tabular}
%\end {ruledtabular}
%\end {table}
\br
\endTable

\section{Analytic model}
\label{Analytic}

\subsection{Ginzburg-Landau master equation}

A hypothesis has been previously stated \cite{Vasil'ev01,Vasil'ev04} %[9,13]
that the SR in a semiconductor is mediated by a formation of the transient
BCS-like state of e-h pairs. In order to interpret the SR emission in terms of BCS-like transition one should transform Eqs.(\ref{TravAmpl}) and (\ref{P&n}) to the form of the Ginzburg-Landau equation (GLE) for the second-order BCS phase transition. The difficulty is caused by the fact the GLE is essentially a steady-state equation for an isolated system, while the SR is a transient process. It appears that to overcome this difficulty we have to change variables to the internal coordinates of the %forward and backward
SR pulse that is, to consider the process in a moving coordinate system. Prior to reporting the technical details of such coordinate transform, it is useful to provide a physical insight into the nature of condensing quasiparticles and the meaning of the coordinate transform into a moving coordinate system.

\begin {figure}[tbp]
\includegraphics{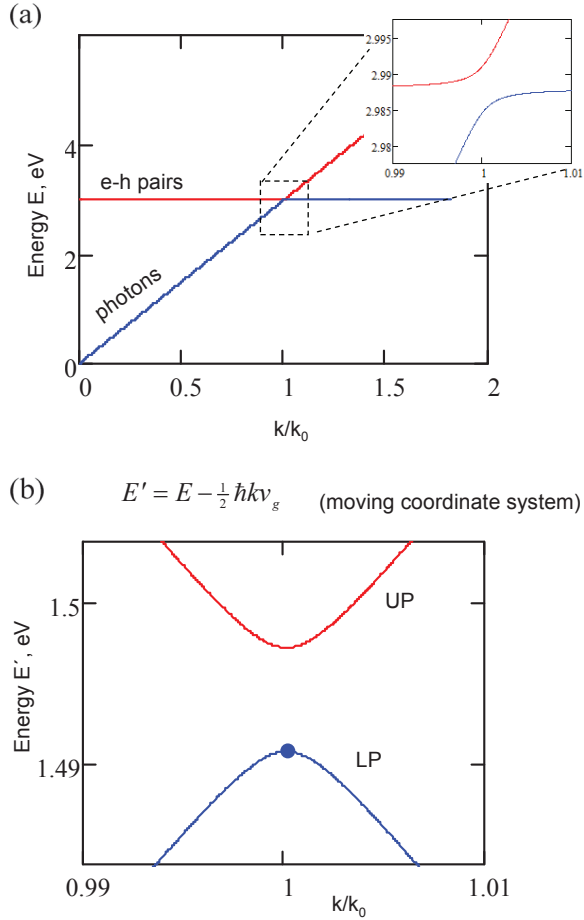} % use for pdf
\caption {(a) Energies of longitudinal 1D polaritons. The inset shows the band anticrossing at the band slope of $v_g/2$. (b) Same energy diagram in the moving at $v_g/2$ coordinate system. The dot indicates the LP´ polariton trap. }\label {fig1_Dispersion}
\end {figure}

Figure \ref{fig1_Dispersion} (a) shows the dispersion of the reduced energy of e-h pair in a direct band gap semiconductor in function of the wavevector component along the cavity axis. (The e-h pair fulfills the energy and momentum conservation under the interband transition with emission of a photon). The e-h pair energy dispersion is plotted alongside with the dispersion of photons in the guided mode of the cavity. The scale is set by the wave number $k_0$ of photons at the interband transition energy and is so small that the dispersion of e-h pair energy $\propto k^2$ is not seen in Fig. \ref{fig1_Dispersion}.
Since we are interested in a small region around $k_0$, the effects of optical waveguide mode cut-off or any other features yielding a deviation from the linear dispersion are not considered here.

The coupling of e-h pairs with photons results in formation of lower polariton (LP) and upper polariton (UP) dispersion branches, which are the eigen solutions of the Jaynes-Cummings model \cite{JCM63,Tavis68}.
We thus focus here on features related to 1D dispersion (and condensation) of such longitudinal polaritons in an edge emitting cavity, while two other degrees of freedom has minor impact on the results reported here.

Jaynes-Cummings model predicts that  LP and UP band anticrossing occurs at $k_0$. For a typical photon number in the SR pulse (see Sec.\ref{Results}), the Jaynes-Cummings model predicts a gap of ~5 meV (see the inset in Fig.\ref{fig1_Dispersion}). The group velocity of UP and LP at the band anticrossing  is $\partial E /\hbar \partial k = v_g/2$, a half of the cavity mode group velocity, indicating that longitudinal 1D polaritons in edge-emitting cavities are the half matter - half light composite bosons. In order to meet the general considerations admitting 1D condensation of polaritons, we shall now identify the energy trap for such longitudinal polaritons.

Figure \ref{fig1_Dispersion} (b) shows the UP and LP band anticrossing plotted in the coordinate system co-propagating with UP and LP at $v_g/2$ velocity. The relativistic correction at such coordinate transform is negligible. Thus $1/\surd ( 1- v_g^2/4c^2) \approx 1.01 $ for parameters shown in Table \ref {Param}. The UP´ (LP´) energies are obtained using the transformation $E^{\prime}=E-\frac 12 \hbar k v_g$ .

In Fig.  \ref{fig1_Dispersion} (b), the band edge of LP´ with negative effective mass represents an energy trap, thus admitting condensation of lower polaritons. The effective mass of LP polaritons $ |m_{LP}|= |\partial E/ \hbar^2 \partial k^2 | \sim 10^{-11}m_0$ assumes possibility of room temperature condensation, during SR pulse emission ($m_0$ is the free electron mass). The thermal de Broglie wavelength criteria for macroscopic quantum degeneracy predicts very low critical densities, indicating that thermodynamic considerations do not provide the main limiting factor for LP´ condensation and SR pulse emission. Therefore it would be reasonable to expect that in the first approximation, the temperature will not enter the GLE for SR emission in semiconductors.

The condensate, as we have shown above, is expected to move at $v_g/2$ velocity.  Technically, it is much more simple to perform analysis in a coordinate system co-moving with the building SR pulse in the cavity, as compare to the system moving  at $v_g/2$. The coordinate transformation to the internal pulse coordinates has received a wide spread use in the analysis of cooperative effects and resonance amplification in masers \cite{Arecchi70,Arecchi65}.  For simplicity, we consider only the forward traveling pulse, introducing  new coordinates  $\zeta{=}t-z/v_g$ and $z$ (instead of $t$ and $z$, respectively). As shown above, in the case of SR in semiconductors, we may ignore the relativistic corrections.

The phase transition under the question is related to the change of the effective phase relaxation time $T_2^{\rm{eff}}$, which is comparable to the lifetime of photons in the cavity  ($1/v_g \alpha_i$=3ps). The decay of macroscopic polarization due to dephasing of e-h pairs cannot be neglected. On the other hand, the carrier injection into the active region (e.g. into QWs)  and carrier relaxation at the rate $1/\tau_n$ can be neglected
%. The material losses also impact the critical density, however   neglect by \textbf{DEL( dephasing and other)} slowly varying terms
at the time scale of the pulse width, yielding a system of equations that is similar to the one used in studies of SR in atomic ensembles
\begin{eqnarray}
%\begin{equation}
%\begin{split}
\frac{\partial A_+}{\partial z}&{=} \frac{\Gamma}{2 v_g} \sqrt{\frac{g_0}{T_2}}
%\frac{\sqrt{g_0 }}{2\sqrt{T_2}}
P_{+} {-}\frac {\alpha_i}2  A_+, \\
\frac{\partial n}{\partial \zeta}&{=}{-}\sqrt{\frac{g_0}{T_2}}A_{+} P_{+}, \\
\frac{\partial P_{+}}{\partial \zeta}&{=}{-}\frac{P_+}{T_2}{+}\sqrt{\frac{g_0}{T_2}}(n{-}n_t)A_+.
\label{Space1}
%\end{split}
%\end{equation}
\end{eqnarray}
Usually, such system of equations has been solved by introducing an instantaneous (partial) pulse area variable, which transforms it into an equation for a non-harmonic pendulum \cite{MacGillivray76,Arecchi70}. The anaclitic solution has been possible so far only for a point source considered by Dicke. In the case of elongated (extended) source, it has  been solved only numerically.

The key to our analytic solution of  this system  was an observation that the optical field of the SR pulse follows the macroscopic polarization of the medium within a certain domain of size $L^*$. This can be seen from the features spotted by the numerical solution in Fig. \ref{fig345_TZDyn}. (Compare Figs. (b) and (d)  as well as (c) and (e); further discussion can be found in Sec. \ref{Results}).
Therefore the analytical solution
is obtained by substituting $1/L^* \rightarrow \partial/ \partial z$ with $L^*$ being the  effective size of the domain with large macroscopic polarization of the medium (condensate fraction), yielding
\begin{equation}
A_+(\zeta,z){=}\sqrt{\frac{g_0}{T_2}} \frac{\Gamma}{v_g(\alpha_i{+}2/L^*)} P_+(\zeta,z).
\label{AtoP}
\end{equation}
This original anzats  allows us to exclude the wave amplitude from Eq.(\ref{Space1}) and convert our problem of SR in a cavity waveguide into a task analytically similar to the well studied case of SR emission from a point source:
\begin{eqnarray}
%\begin{equation}
%\begin{split}
\frac{\partial P_+}{\partial \zeta}&{=}\frac{\Gamma g_0 (n{-}n_{cr}^*)}{ T_2 v_g (\alpha_i {+} 2{/}L^*) }P_+, \\
\frac{\partial n}{\partial \zeta}&{=}
%{-}\frac{\Gamma g_0 P_+^2}{v_g T_2 (\alpha_i{+}2/L^*)}{=}
{-}\frac{P_+}{(n{-}n_{cr}^*)}\frac{\partial P_+}{\partial \zeta},
%\quad \frac{\partial A_+}{\partial z}{=}\frac 12 \Gamma\ \tilde{P}_+ -\frac 12 v_g \alpha_i A_+,
%\\
\label{Space2}
%\end{split}
%\end{equation}
\end{eqnarray}
where we have introduced a new parameter
\begin{equation}
n_{cr}^*{=}n_t{+}\frac{v_g(\alpha_i{+}2/L^*)}{\Gamma g_0},
\label{Ncr}
\end{equation}
which, as shown later, is the critical density for the condensation phase transition.
Following recipes established for a point source, the last equation in (\ref{Space2}) allows one to recover the Bloch vector conservation in a cavity waveguide
\begin{equation}
(n(\zeta,z){-}n_{cr}^*)^2{+}P_{+}^2(\zeta,z){=}(n_0{-}n_{cr}^*)^2
\label{IntMotion}
\end{equation}
with $ n_0{=n}({-} \infty)$ being the initial carrier density in the system, before the emission of SR pulse.

However, taking the second derivative in the first equation (\ref{Space2}) and  substituting $\partial n / \partial \zeta$ and $\partial P_+ / \partial \zeta$ from  (\ref{Space2}), we obtain the following  master equation for SR emission and polariton condensation in an elongated semiconductor sample (cavity waveguide):
\begin{equation}
%\begin{split}
%\hspace{-0.01in}
{-}\frac{ \partial^2 P_+}{\partial \zeta^2 }{+}
\frac{\Gamma^2 g_0^2 {L^*}^2 }{T_2^2 v_g^2(2{+}\alpha_i L^*)^2}\Bigl[(n_0{-}n_{cr}^*)^2
{-} 2 P_+^2 \Bigr]P_+{=}0.
\label{GLE}
%\end{split}
\end{equation}

This equation has a canonical form of the Ginzburg-Landau equation (GLE) (or Gross-Pitaevskii equation), in which     $P_+$ is the order parameter of the system and the internal coordinate $\zeta{=}t{-}z/v_g$ plays the role of spatial variable. For the carrier densities above  critical $n_{cr}^*$ (see Eq.(\ref{Space2})), its steady state solution defines the condensate fraction. The analytical similarities between Eq.(\ref{GLE}) and GLE, allow one to assume that for $n{>}n_{cr}^*$, the evolution of the order parameter is defined by
hyperbolic secant function:
\begin{equation}
P_+(\zeta,z){=}P_0\rm{sech}(\zeta/\tau_p)
\label{GLEsolution}
\end{equation}
Substituting  in (\ref{GLE}), %$\tilde{P}_+{=}\tilde{P_0}/\cosh(z/\tau_p) $
%$P_+{=}P_0\text{sech}(\zeta/\tau_p)$,
we obtain the parameters $P_0$ and $\tau_p$:
\begin{eqnarray}
%\begin{equation}
%\begin{split}
P_0&=n_0-n_{cr}^*,\\
\tau_p&{=}\frac{ T_2 v_g {(}\alpha_i {+} 2/L^*{)}}{\Gamma g_0 (n_0{-}n_{cr}^*)}.
\label{Tcoh}
%\end{split}
%\end{equation}
\end{eqnarray}
At the peak of SR pulse, the order parameter $P$ thus reaches the highest value  admitted by the integral of motion (\ref{IntMotion}).

Using normalization convention (\ref{NormA}) and relationship (\ref{AtoP})
we find the evolution of photon number in the cavity $A_{+}^2(\zeta,z)$ and
%the  photon number
at the peak of SR  pulse
\begin{eqnarray}
%\begin{equation}
%\begin{split}
%\tau_p&{=}\frac{ T_2 v_g {(}\alpha_i {+} 2/L^*{)}}{\Gamma g_0 (n_0{-}n_{cr}^*)},\\
&{A_+^2(t{-}z{/}v_g)}{=}\frac{T_2}{g_0\tau_p^2} \frac 1 {\rm{cosh}^2(\frac{t-z/v_g}{\tau_p})}\\
&N_{+}{=}A_{+}^2(0){=}
%A_{+}^2(0){=}
\frac{T_2}{g_0\tau_p^2} %\propto(n_0{-}n_{cr})^2
{=}\frac{\Gamma^2 g_0(n_0-n_{cr}^*)^2}{T_2 v_g^2 (\alpha_i +2/L^*)^2}
\label{TcohN}
%\end{split}
%\end{equation}
\end{eqnarray}
For convenience and in order to shorten our formulas, we use the following convention for the pulse width and coherence time \cite{CohExpl}
\begin{equation}
\tau_c=2\tau_p
\label{TcohDef}
\end{equation}
The pulse width %$\tau_c=2\ln(1{+}\sqrt{2})\tau_p$
$\tau_c \propto 1/(n_0-n_{cr}^*)$
and the peak power density
$ \hbar \omega v_g N_{+} \propto (n_0-n_{cr}^*)^2$
  exhibit the expected behavior
known from the studies of SR in an atomic ensemble % gas medium
 \cite{Dicke54,Andreev93}.

We thus have obtained that SR in a semiconductor follows the GLE (\ref{GLE}) for the
 phase transition to macroscopic coherent state and the medium polarization $P_{\pm}$ is the order parameter of the system.
The proposed consideration of
condensation during SR emission is different from all previous treatments. It does not involve the system temperature. As shown above,
the effective mass of condensing quasiparticles (1D lower polaritons) is so small ($\sim 10^{-11}m_0$)  that the critical density is very low. This renders the approximation (\ref{GLE}) to be valid even though the thermodynamic temperature of the system is not accounted for. The thermal de Broglie wavelength criteria for macroscopic quantum degeneracy is satisfied at the carrier densities much below the one stated in Eq.(\ref{Ncr}).

\subsection{Type-I superradiance regime}
\label{SecSR1}
The features of the SR pulse %expressed in
(\ref{TcohN})
are in  perfect agrement with predictions of the Dicke theory if the coherence length $L_c{=}\tau_c v_g$ (as measured in the cavity) is longer than the sample size $L$ (the cavity length \cite{SampleSize}). In this regime, refereed herewith as the type-I SR, the characteristic length of the condensate fraction $L^*$ is limited by the sample size  $L^*{=}L{<}L_c$. From Eqs. (\ref{Ncr}), (\ref{Tcoh}) and (\ref{TcohN}), we obtain the critical carrier density $n_{cr}^{(I)}$, pulse width parameter $\tau_p^{(I)}$ and the peak photon number $N_+^{(I)}$
\begin{eqnarray}
%\begin{equation}
%\begin{split}
n_{cr}^{(II)}&>n_0>n_{cr}^{(I)}{=}n_t{+}\frac{v_g(\alpha_i{+}2/L)}{\Gamma g_0},\\
\tau_p^{(I)}&{=}\frac{ T_2 v_g {(}\alpha_i {+} 2/L{)}}{\Gamma g_0 (n_0{-}n_{cr}^{(I)})},\\
N_{+}^{(I)}&{=}
%A_{+}^2(0){=}
\frac{T_2}{g_0(\tau_p^{(I)})^2} %\propto(n_0{-}n_{cr})^2
{=}\frac{\Gamma^2 g_0(n_0-n_{cr}^{(I)})^2}{T_2 v_g^2 (\alpha_i +2/L)^2},
\label{Tcoh1}
%\end{split}
%\end{equation}
\end{eqnarray}
The second critical density $n_{cr}^{(II)}$ introduced here is defined in the next  Section \ref{SecTypeII}.

The output pulse power can be obtained using the photon density in the cavity
(\ref{TcohN}) and accounting for the reflection of the laser chip facets:
\begin{eqnarray}
%\begin{equation}
%\begin{split}
%N_+(\zeta,z)&{=}A_+(\zeta,z)^2{=}\frac{T_2}{g_0}
%\frac{1}{{\tau_p^2 }\text{cosh}^2(\zeta/\tau_p)}\\
P_{out}^{(I)}(z,t)&{=}\hbar \omega(1-R)v_g
%\frac {w d_{QW} N_{QW} }{\Gamma}
\Omega
%\frac{T_2}{g_0 \tau_p^2 }
\frac{N_+^{(I)}}{\rm{cosh}^2(\frac{t-t_0-z/c}{\tau_p^{(I)}})}
%\\
%&{=}\hbar \omega(1-R)v_g
%\Omega
%\frac{\Gamma^2 g_0(n_0-n_{cr}^{(I)})^2}{T_2 v_g^2 (\alpha_i +2/L)^2} \\
%&{\times} \frac{1}{\text{cosh}^2(\frac{t-t_0-z/c}{\tau_p})}
%\end{split}
\label{IntensitySR1}
%\end{equation}
\end{eqnarray}
where $R$  is the reflection coefficient of the facet, $\Omega = w d_{QW} N_{QW} / \Gamma$ is the mode cross-section area,  $w$ is the waveguide width, $d_{QW}$ and $N_{QW}$ are the thickness and number of quantum wells and $\Gamma$ being the optical confinement factor. The pulse envelope in Eq.(\ref{IntensitySR1}) accounts for free-space propagation of the pulse, and $t_0$ is the delay to emission of the SR pulse. Note that within the semiclassical approach  we use here, the time to emission of the pulse cannot be defined and requires quantum-mechanical consideration \cite{Dicke54,Andreev93}.

\subsection{Type-II superradiance regime}
\label{SecTypeII}
With increasing carrier density, the coherence length $L_c=2\tau_p v_g$ reduces.
At $n_0>n_{cr}^{(II)}$, it becomes shorter than the sample length $L$
indicating that the size of the domain, which is occupied by the condensate fraction, is % set by the coherence length
smaller than the sample size ($L^*{=}L_c{<}L$). We refer this regime as the type-II SR. As shown below, in distinguishing  from type-I SR regime, the pulse peak power does not exhibit  quadratic growth
$\propto (n_0-n_{cr})^2$. In principle, this is the only SR regime that can be observed in experiments with  edge-emitting semiconductor laser cavities of several  hundreds of microns length
(see Sec. \ref{Results}).

The critical density $n_{cr}^{(II)}$
can be obtained from Eq. (\ref{Tcoh1}). At this carrier density,
 the coherence length coincides with the sample size so as $\tau_c=L /v_g$. Substituting
 $\tau_c/2$  in the l.h.s. of  %expression
 (\ref{Tcoh1})
  and solving it with respect to $n_0$, we find the the critical carrier density for the type-II SR regime is
 \begin{equation}
%\hspace{-0.001in}
%n_0{>}
n_{cr}^{(II)}{=}n_t{+}\frac{v_g(2{+}\alpha_i L)}{\Gamma g_0 L}{\Bigl(}1{+}\frac {2
%\ln(1{+}\sqrt2)
T_2 v_g}{L} {\Bigr)}>n_{cr}^{(I)}{.}
{\label{Ncr2}}
\end{equation}

At $n_0>n_{cr}^{(II)} $, the effective size of the condensate fraction domain $L^*$ in (\ref{Ncr}) and  (\ref{Tcoh}) % is smaller than the cavity sizy and
is conditioned  by the coherence length %time itself,
$L^*{=}L_c{=}2 v_g  \tau_p^{(II)} $ %\ln(1+\sqrt2)$
and we obtain a quadratic equation with respect to $\tau_p^{(II)}$ that reads
\begin{equation}
\tau_p^{(II)}{=}\frac{ T_2 {(}v_g\alpha_i {+} 1/\tau_p^{(II)}{)}}{\Gamma g_0 (n_0{-}n_t){-}(v_g\alpha_i{+}1/\tau_p^{(II)} )}
\end{equation}
This equation has only one positive root, yielding
\begin{eqnarray}
%\begin{equation}
%\begin{split}
\tau_p^{(II)}&{=}\sqrt{\frac {T_2}{\Gamma g_0 (n_0{-}n_t)}}\\
{\times}&\Biggl(\sqrt{1{+}\frac{(1{-}\alpha_i v_g T_2)^2}{4 \Gamma g_0 T_2 (n_0{-}n_t)}} {-}\frac{1{+}\alpha_i v_g T_2}{2\sqrt{\Gamma g_0 T_2 (n_0{-}n_t)}}
\Biggr)^{-1}{.}
%\\
%N_+ &{=}\frac{T_2}{g_0\tau_p^2}
\label{Tcoh2}
%\end{split}
%\end{equation}
\end{eqnarray}

In the type-II SR regime, the effective size of the condensate fraction domain is smaller than the sample size  $L_c=2 v_g \tau_p^{(II)} <L$. At each moment of time, this domain is centered at the peak of the growing SR pulse. However, considering the build-up of SR pulse propagating through the sample, we shall split our sample into domains of size $L_c$ and sum up the contributions from all domains in the sample. In the laboratory coordinate system $(t,z)$, we thus have to account for the fact that different sample domains do not contribute to the SR pulse simultaneously.

However, these calculations are particularly straightforward in the internal coordinate system of the pulse $(\zeta,z)$ that we use here.
 We have to account only for incoherent superposition of wavelets emitted from different domains in the sample. Otherwise, if the domains are mutually coherent, the coherence length $L_c$ will not be smaller than the sample size $L$, leading to type-I SR regime.

There are thus ${\sim}L/L_c$ mutually incoherent domains. % wavelets in total.
We introduce index $j$ to enumerate these domains and corresponding radiated wavelets $A_{+,j}$.
For the order parameter of each domain (polarization) %at $z^{(j)}$
 and slowly-varying amplitude of the radiated wavelets,
we use Eqs. (\ref{GLEsolution})-(\ref{TcohN}):
\begin{eqnarray}
%\begin{equation}
%\begin{split}
P_{+,j}&{=}\frac {n_0-n_{cr}^{(II)}}{\rm{cosh}(\zeta /\tau_p^{(II)})}e^{i \phi_j}\\
%=\frac { (\alpha_i  + 2/L_c)v_g T_2} {\tau_p \Gamma g_0}\frac{1}{\text{cosh}(\zeta /\tau_p)}\\
A_{+,j}&{=}\sqrt{\frac{T_2}{g_0}}
\frac{e^{i \phi_j}}{{\tau_p^{(II)} }\rm{cosh}(\zeta/\tau_p^{(II)})}.
%\end{split}
\label{P_SR2}
%\end{equation}
\end{eqnarray}
The random phases $\phi_j$ account for the mutual incoherences between domains of condensate fraction. The corresponding superposition of wavelets reads
\begin{eqnarray}
%\begin{equation}
%\begin{split}
%P_{+,j}&{=}\frac {P_0}{\text{cosh}(\zeta /\tau_p)}
%=\frac { (\alpha_i  + 2/L_c)v_g T_2} {\tau_p \Gamma g_0}\frac{1}{\text{cosh}(\zeta /\tau_p)}\\
A_+(\zeta,z)=\sum_j A_{+,j} e^{-\frac 12 \alpha_i (z-z^{(j)})}\Theta(z-z^{(j)})
%&{=}\sum_j \sqrt{\frac{T_2}{g_0}}
%\frac{1}{{\tau_p^{(II)} }\text{cosh}(\zeta/\tau_p^{(II)})}
,
%\end{split}
\label{A_SR2}
%\end{equation}
\end{eqnarray}
where the exponential factor accounts for material loss and $\Theta(z)$ is the Heaviside step function, which accounts for the fact that a wavelet emitted at some $z^{(j)}>z$ and propagating in the positive z-axis direction does  not contribute to the field at $z<z^{(j)}$.

The superposition of incoherent wavelets (\ref{P_SR2}) defines the evolution of photon number in the cavity $N_+(\zeta,z) = \langle | A_+(\zeta,z)| ^2 \rangle $. In particular, the photon density at the %right
cavity facet reads
\begin{eqnarray}
%\begin{equation}
%\begin{split}
N_+(\zeta,L)&{=}\Bigl\langle \Bigl | \sum_j A_{+,j}e^{-\frac 12 \alpha_i (L-z^{(j)}) } \Bigr | ^2 \Bigr\rangle \\
&{=}\sum_j \Bigl \langle \Bigl | A_{+,j}\Bigr |^2 \Bigr \rangle e^{-\alpha_i (L-z^{(j)}) }\\
&{=}\frac{T_2}{g_0 (\tau_p^{(II)})^2 }
\frac{1-e^{-\alpha_i L}}{{1-e^{-\alpha_i L_c}}}\frac{1}{\rm{cosh}^2(\zeta/\tau_p^{(II)})}.
%\end{split}
%\label{N_SR2_simp}
\label{N_SR2_exact}
%\end{equation}
\end{eqnarray}
Here the ensemble average is taken over the mutually incoherent domains of the sample. The cross-correlation terms between $i$-th and $j$-th domains vanish since the relative phase $\phi_i-\phi_j$ is random. Thus the factor $(1-e^{-\alpha_i L})/(1-e^{-\alpha_i L_c})$ distinguishes the type-II SR from its type I counterpart (compare with Eq.(\ref{Tcoh1})).

The peak photon number and the output pulse power in the type-II SR regime are  \begin{eqnarray}
%\begin{equation}
%\begin{split}
N_+^{(II)}&{=}\frac{T_2}{g_0 (\tau_p^{(II)})^2 }
\frac{1-e^{-\alpha_i L}}{{1-e^{-2\alpha_i \tau_p^{(II)} v_g }}}
%\frac{1}{\text{cosh}^2(\zeta/\tau_p^{(II)})}
,
\\
%&{=}\frac{T_2}{g_0\tau_p^2}
%\frac { (2+\alpha_i L_c )} {\alpha_i L_c}(1-e^{-\alpha_i L })\\
P_{out}^{(II)}(z,t)&{=}\hbar \omega(1-R)v_g
\frac{N_+^{(II)}}{\rm{cosh}^2(\frac{t-t_0-z/c}{\tau_p^{(I)}})}
%{\propto}(n_0{-}n_{cr})
%\frac{\Gamma^2 g_0(n_0-n_{cr})^2}{T_2 v_g^2 (\alpha_i +2/L^*)^2}
\label{IntensitySR2}
%\end{split}
%\end{equation}
\end{eqnarray}
where we have used that  $L_c=2v_g \tau_p^{(II)} $.

At large carrier densities $n_0\gg n^{(II)}_{cr}$, the pulse width (coherence time) reduces, yielding the asymptotic behavior
\begin{eqnarray}
%\begin{equation}
%\begin{split}
\tau_p^{(II)}&{\sim} \sqrt{\frac {T_2}{\Gamma g_0 (n_0{-}n_t)}},
\\
N_+^{(II)} & {\sim}
\frac{T_2(1-e^{-\alpha_i L})}{2 g_0 \alpha_i v_g (\tau_p^{(II)})^3 }
\\
&{=}
\frac{\Gamma^{3/2}}{2 \alpha_i v_g} \sqrt{\frac{g_0}{T_2}}(n_0{-}n_{t})^{3/2}(1-e^{-\alpha_i L})
%\frac{\Gamma^2 g_0(n_0-n_{cr})^2}{T_2 v_g^2 (\alpha_i +2/L^*)^2}
\label{Tcoh2_ass}
%\end{split}
%\end{equation}
\end{eqnarray}
Thus in the type-II SR regime, because only partial coherence can be achieved over the sample ($L_c<L$), the photon number does not follow the  quadratic growth predicted by Dicke' model.  At the same time, the growth of $N_+^{(II)}$ is superlinear,
%in agreement with the fact of partial coherence (
as opposed to the linear growth in an incoherent ensemble.  The pulse energy $\propto N_+^{(II)}\tau_p^{(II)}$ is proportional to $(n_0-n_t)$, in agreement with the energy conservation considerations.

With increasing sample length $L$, the difference between the first $n_{cr}^{(I)}$ (\ref{Tcoh1}) and  second $n_{cr}^{(II)}$ (\ref{Ncr2}) critical densities reduces, indicating that the cavity tends to operate in the type-II SR regime provided special provisions are  taken to prevent lasing. At very large sample length, both critical densities are clamped at $n_t+v_g \alpha_i/\Gamma g_0$. It can be seen that some excess above the transparency carrier density is always required to  compensate for material losses.

If the optical losses are negligible ($\alpha_i \ll 1/L$), the photon  number and the output peak power in the type-I  SR regime increase as ${\propto} L^2$
provided the conditions of the type-I  SR regime are fulfilled, in particular that $L^2{<}4T_2v_g^2/\Gamma g_0 (n_0-n_t)$.

In the lossless limit of the  type-II SR regime, the photon number and the output pulse peak power scale as  $\propto L$ with the sample size. At same time if some  losses are present in the cavity, the photon number and the peak power show a saturation, starting from the cavity length $L\sim 1/\alpha_i$.

\section{Results and discussion}
\label{Results}

\begin {figure}[tbp]
\includegraphics [width=7.1cm]{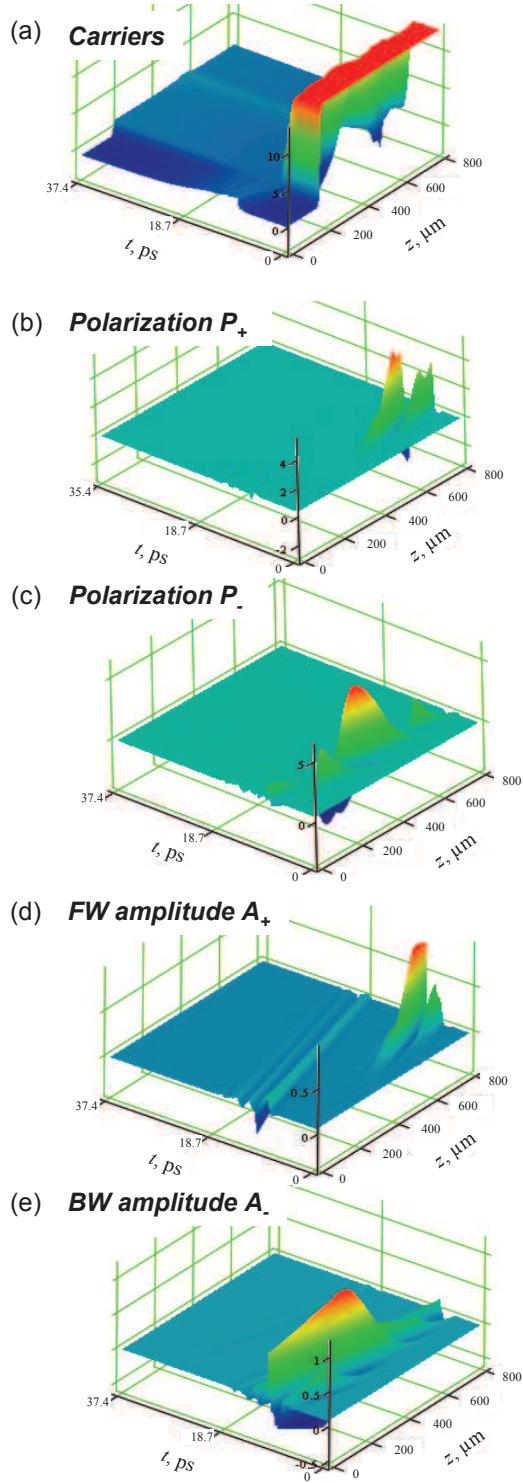}
\caption {Numeric simulations: Spatiotemporal dynamics of the carrier population (a), macroscopic polarization in InGaN/GaN QWs associated with the forward FW (b) and backward BW (c) waves; field amplitudes of the FW  (d) and BW (e) waves. The overall cavity length is 800 $\mu$m ($z$ axis). The absorber of 160 $\mu$m long is situated at the beginning of the coordinates. The time interval of two cavity roundtrips is shown ($t$ axis). All parameters are normalized on carrier density at transparency  $n_t$.
The current amplitude is $I$ = 281 mA (the current density $J$=22kA/cm$^2$).
}
\label {fig345_TZDyn}
\end {figure}

\begin {figure}[tbp]
\includegraphics {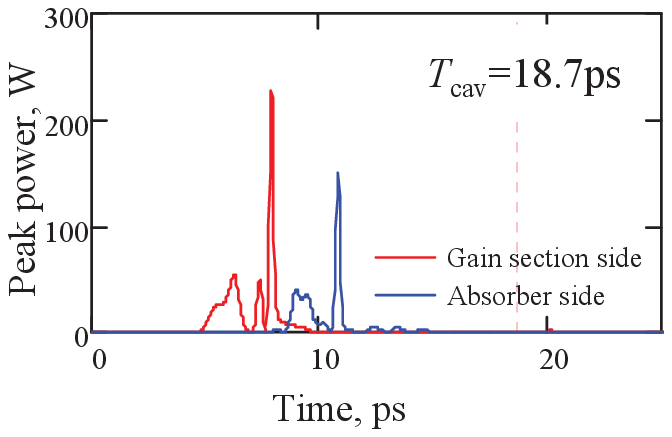} % use for Tex
\caption {Numerical simulations: Output SR pulses at the gain section facet (red curve) and absorber section facet (blue curve). The cavity length is 800 $\mu$m, the absorber length is 160 $\mu$m and driving conditions are given in the caption of Fig. \ref{fig345_TZDyn}.}\label {fig3_Output}
\end {figure}

The validity of analytical model predictions
is confirmed
by numerical simulations in an edge-emitting laser cavity incorporating InGaN/GaN double quantum well heterostructure (Table \ref{Param}).
The numeric model used here is based on the traveling wave Maxwell-Bloch equations of Sec. \ref{num_mod}. It reproduces spontaneous triggering of the SR emission by  incorporation of the spontaneous polarization term $\Lambda_\pm$ in Eq.(\ref{P&n}). We also implement the boundary conditions  at the left ($z{ =}0$) and right ($z {=} L$) cavity facets in order to mimic the effect of accumulation of spontaneous photons in the cavity before the SR pulse emission and in order to analyze possible effects of multiple cavity roundtrips due to reflections of the cavity facets
\begin{equation}
%\begin{split}
A_{+}(0,t){=}\sqrt R A_{-}(0,t),\quad A_{-}(L,t){=}\sqrt R A_{+}(L,t).
%\end{split}
\label{BC}
\end{equation}
We complete the numerical model by incorporating separate equations for a possible absorber section in the cavity. This enables us to compare the analytic model predictions for the SR pules and condensation of longitudinal polaritons (in the gain section) with the numerical simulations for the entire cavity representing the actual  experimental arrangement.

 The model equations for the carrier population and polarization dynamics in the absorber section take into account (i) the Quantum Confined Stark Effect (QCSE) energy shift of the absorption edge
and (ii) the reduced carrier lifetime in a negatively biased absorber:
\begin{eqnarray}
%\begin{equation}
%\hspace{-0.1in}\begin{split}
&\frac{\partial P_{\pm,a}}{\partial t }{=}{-}\frac{P_{\pm,a}}{T_2}
%{+}D\frac{\partial^2 P_\pm}{\partial z^2}
%{+}\sqrt{\frac{g_0}{T_2}}(n{-}n_t) A_\pm
%\\ &\quad\quad
{+}\sigma\sqrt{\frac{g_0}{T_2}}(n_a{-}n_V) A_\pm
%{+} \Lambda_\pm
%\\
%&\frac{\partial n}{\partial t }{=}
%%D\frac{\partial^2 n_\pm}{\partial z^2}
%{-}\frac{n}{\tau_n}{-}\sqrt{\frac{g_0}{T_2}}(A_{+}P_{+}{+}A_{-}P_{-}){+}\frac{J(z{,}t)}{e_q d}
\\
&\frac{\partial n_a}{\partial t }{=}
%D\frac{\partial^2 n_a}{\partial z^2}
{-}\frac{n_a}{\tau_a}{-}\sigma\sqrt{\frac{g_0}{T_2}}(A_{+}P_{+,a}{+}A_{-}P_{-,a})
\label{P&n_Abs}
%\end{split}
%\end{equation}
\end{eqnarray}
where
$\sigma$  is the differential absorption to gain ratio, $\tau_a$ is the carrier lifetime in absorber section. The parameter $n_V$ accounts for the QCSE energy shift
 in the absorber section QWs caused by the external negative bias.
 In our numeric model simulations, we use $\sigma=3.5$ and $\tau_a=5$ ps while the normalized absorber bias $V_a{=}(n_t{-}n_V)/n_t$ is in the range form 0 to -1 (highly absorbing state of absorber).

Figure \ref{fig345_TZDyn} displays the results of numerical simulations
for the SR emission in the cavity of the overall length $L$=800$\mu$m incorporating a 160 $\mu$m long absorber. The absorber section is located at the left facet of the structure (at $z{=}0$) and the gain section is situated at the right facet.
The spatiotemporal dynamics of the carrier population (Fig.\ref{fig345_TZDyn} (a)), order parameter  (Figs.\ref{fig345_TZDyn}(b),(c)) and the optical field (Figs. \ref{fig345_TZDyn} (d),(e)) as well as the intensities of the output SR pulses  emitted at both facets of the structure (Fig.\ref{fig3_Output}) are quite complicated.

Initially, a high reversed bias is applied to the absorber leading to depletion of carriers in the QWs  (blue color in the scale of Fig.\ref{fig345_TZDyn}(a)
). It prevents the structure from lasing and enables accumulation of carriers in the gain section. The gain section QWs are pumped to a high level, so as in this particular case, the carrier population is about 15 times above the transparency $n_t$ (red color in Fig.\ref{fig345_TZDyn}(a). The entire process starts at $t$=0 from the spontaneous polarization noise.
The rate of spontaneous polarization noise into the cavity mode ${\sim} \frac 1 2 \Gamma \sqrt{g_0 T_2}  \Lambda_{\pm}$
corresponds to
27 photons per cavity round trip.
This cannot be seen in the scale of Fig.\ref{fig345_TZDyn}.
The noise sources are uniformly distributed along the cavity %(Z axis)
but the macroscopic coherences  are building up
at the
edges of the gain section,
after about half of the cavity roundtrip time ($T_{\rm{cav}} $ =18.7ps). Because of the asymmetric cavity,
the shape and emission time of the output SR pulses from the left and right cavity facets are different (Fig.\ref{fig3_Output}) and the emitted SR pulses
are not related by reflections at the cavity facets.

The spatio-temporal dynamics of the order parameter $P_{\pm}$ shown in Figs. \ref{fig345_TZDyn} (b) and (c) clearly indicates that during all process of SR, the coherence length remains shorter than the gain section length ($L_c{<}L_s$).
Interestingly, in Fig.\ref{fig345_TZDyn} (e),
the peak intensity of the pulse
corresponds to photon density in the cavity of 1.4$\cdot$10$^{19}$ cm$^{-3}$, while the density of the associated coherence excitations $P_{\pm}$ at the polarization peak maxima is of 9.9$\cdot$10$^{19}$ cm$^{-3}$, which is of the order of
the initial carrier density $n_0$=2.1$\cdot$10$^{20}$  cm$^{-3}$ .
Thus the associated photon density in the cavity is, at first sight, quite low.
In fact the dynamic variables in Eq. (\ref{TravAmpl}) account for the optical confinement factor $\Gamma$ due to the difference between the cavity mode size  and the QW width. The effective peak photon density reduced to the QWs width is of 7$\cdot$10$^{20}$ cm$^{-3}$, which is thus spectacularly large as compared with the initial density $n_0$ of uniformly distributed carriers.
In this example, illustrating a typical situation in the edge-emitting  laser cavity, the system operates %thus
in the type-II SR regime, in which the condensate fraction is confined to a domain smaller than the size of the sample.

The build-up of the macroscopic polarization is followed by Rabi oscillations at a frequency dependent on the pumping rate. They are clearly seen in the
electric field amplitudes of the waves showing characteristic oscillations with the reversal of the sign (Figs.\ref{fig345_TZDyn}(d) and (e)). The carrier density drops abruptly at the same time when the polarization rises, indicating that almost all carriers contribute to the field (Figs.\ref{fig345_TZDyn}(a)).  When the travelling backward pulse hits the absorber section at $z{=}160\mu$m, the pulse intensity is sufficiently high to saturate  absorber %and to make it transparent
(the blue step in the region $0{<}z{<}160\mu$m and $t>9$ps in Fig.\ref{fig345_TZDyn}(a)). From this time, there is no absorption and gain in the cavity. The backward travelling SR pulse reflected of the left cavity facet (at $z$=0) can now freely travel through the structure yielding emission of the secondary pulse at the right cavity facet at $t\approx19ps$. Its amplitude can be appreciated from Fig.\ref{fig345_TZDyn} (d). (In the scale of Fig.\ref{fig3_Output}, the intensity of this secondary output pulse cannot be seen.)  In this particular realization of SR emission, the pulse emitted from the absorber section side is of smaller peak intensity.

\begin {figure}[tbp]
\includegraphics {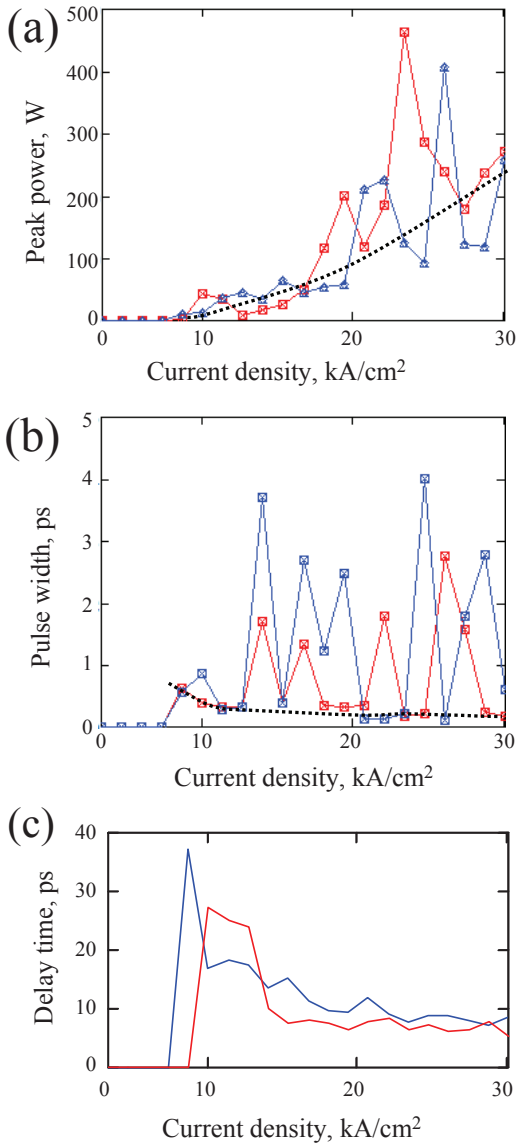}
\caption {Numerical simulations: Peak power (a), FWHM pulse width (b) and delay time (c) of SR pulses emitted at the absorber section facet (blue data points and curves) and gain section facet (red data points and curves) in function of the pump current density. In (a) and (b), we introduce the eye guides (doted black curves) to simplify interpretation of the numerical simulations subjected to large stochastic variations. The cavity length is 800 $\mu$m, the absorber length is 160 $\mu$m.
}\label {fig67_LI}
\end {figure}

The peak power, FWHM pulsewidth of SR pulses and delay time to their emission vary within the range of 50-200 W , 0.4-5 ps and 9-36 ps , depending on the driving conditions
(Fig.\ref{fig67_LI}).
Due to spontaneous polarization
$\Lambda_{\pm}$ in Eq.(\ref{P&n}), the build-up of macroscopic polarization has a stochastic nature, which results in differences between individual realizations of  SR pulses.
In addition, large stochastic variations  of the SR pulse FWHM width between adjacent data points in Fig.\ref{fig67_LI} (b) are caused by the pulse ringing effect, when one of the satellite pulses, as in Fig. \ref{fig3_Output}, reaches the half of the maximum of the main SR pulse.  Therefore the lower boundary of data points in Fig.\ref{fig67_LI} (b) should be regarded as an estimated width of the solitary SR pulse. For convince, we show the eye guides in Figs. \ref{fig67_LI} (a) and (b) [black doted curves].

The time to emission (delay-time) % variations %and jitter
is plotted in Fig.\ref{fig67_LI} (c) as a function of the pump current density. At low pump rates, it takes around 1.5 cavity roundtrip until the emission of the SR pulse and condensation of lower polaritons. The SR pulses at two cavity facets are emitted almost simultaneously. At higher pump current, the SR pulses are emitted after approximately 0.5 cavity roundtrip, following the scenario depicted in Fig.\ref{fig345_TZDyn}.  Note that higher pump levels shorten the time to emission of the SR pulses.

Varying the overall cavity length and the relative length of absorber, we find that the intensity ratio and the width %ratio
of the forward and backward SR pulses can be effectively altered.
The common feature of all realizations
is that the polarization and carrier excess above the transparency
vanish
after a few oscillations so as the
the optical pulses
travel freely
without gain or absorption. This process is accompanied by building up of macroscopic coherent domain in the sample, which we attribute to condensation of 1D longitudinal lower polaritons.

Let us compare the results of numerical simulations with predictions of  simplified analytical model based on solution of the GLE (\ref{GLE}).
In conditions of Fig.\ref{fig345_TZDyn}, the sample size is defined by the the length of the gain section (640 $\mu$m). The critical carrier density of $n^{(I)}_{cr}{=}2.48n_t$ is achieved at the pump current density of 4.17 kA/cm$^2$  ($n_0{=}\tau_n J/e_q d$) so as the type-I condensation condition $n_0{>} n^{(I)}_{cr}$  is fulfilled. However at $n^{(II)}_{cr}=2.52n_t$  (at $J$=4.24kA/cm$^2$) the coherence length reduces down to the size of the cavity. Thus only in the narrow range $n^{(I)}_{cr}<n_0<n^{(II)}_{cr}$, the SR pulse peak power exhibits quadratic growth (\ref{Tcoh1}). At high injection levels, the peak power is $\propto (n_0-n_t)^{3/2}$ [see the asymptotic expression (\ref{N_SR2_exact})]. This feature is perfectly reproduced by the numerical model in Fig. \ref{fig67_LI}.

\begin {figure}[tbp]
\includegraphics {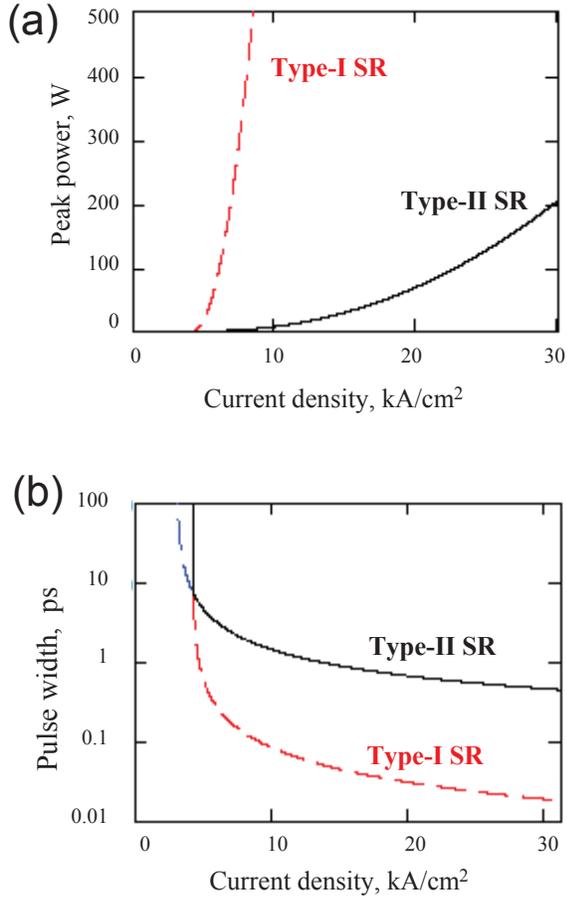} % use for Tex
\caption { Analytical solution of GLE (\ref{GLE}): Peak power (a) and pulse width $\tau_c$ at 42\% of the  peak power (b) of the output SR pulses in type-I [Eq.(\ref{Tcoh1})] and type-II [Eq.\ref{Tcoh2}] regimes in function of the initial carrier density $n_0$. The carrier densities are shown normalized on the transparency carrier density $n_t$. The sample size is 640 $\mu$m , corresponding to the gain section lengths of the cavity used in numerical simulations in Figs. \ref{fig345_TZDyn}, \ref{fig3_Output} and \ref{fig67_LI}. }\label {fig7_GLE_solve}
\end {figure}

Fig. \ref{fig7_GLE_solve} shows the analytical model predictions for the SR output power and pulse width. Although the SR pulse power and pulse width calculated from the numerical model in Fig.\ref{fig67_LI} exhibit large amplitude and timing instabilities,
they are in agreement with the predictions of analytical model.
Thus at the current density 30 $kA/cm^2$, both models predict the SR pulse peak power of $\sim$200 W. The pulse width $2\tau_p$ of 450ps predicted by analytic model (Fig. \ref{fig7_GLE_solve} (b)) corresponds to the SR pulse FWHM $1.76 \tau_p$ of 400 fs \cite{CohExpl}. This nearly reproduces the lower boundary estimates for the pulsewidth obtained from the  numerical model simulations in Fig.\ref{fig67_LI} (b) [see the eye guiding curve].

\begin {figure}[tbp]
\includegraphics [width=8cm]{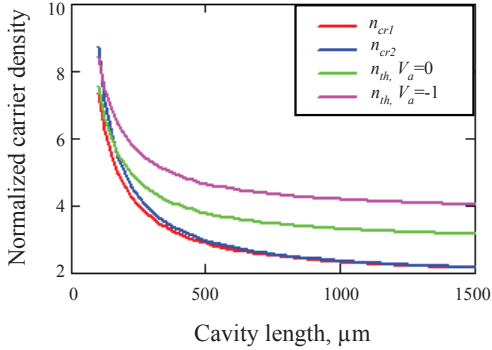} % use for Tex
\caption {Critical carriers densities $n^{(I)}_{cr}$ and $n^{(II)}_{cr2}$ vs cavity length in comparison with the CW lasing threshold carrier densities for the cavity with relative absorber length 20\% in non biased ($V_a=0$) and highly biased ($V_a=-1$) states. The carrier densities are shown normalized on the transparency carrier density $n_t$.}\label {fig6_Ncr}
\end {figure}

Figure \ref{fig6_Ncr} illustrates a relationship between the critical densities for type-I and type-II SR as a function of the overall cavity length $L_{cav}$, assuming that  absorber occupies  20\% of the cavity ($L=0.8L_{cav}$). For comparison, the threshold carrier densities in the CW regime are plotted at lowest ($V_a=0$) and highest ($V_a=-1$) absorption states of the absorber.
 In most of the cases, the type-II SR regime with condensate fraction occupying only a part of the cavity will be reached provided the cavity is made sufficiently long to prevent lasing.

\begin {figure}[tbp]
\includegraphics [width=8cm] {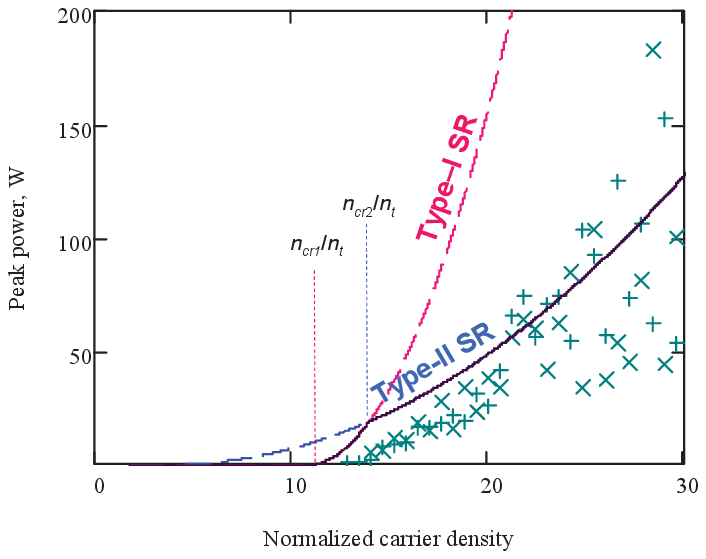} % use for Tex
\caption {Peak power of SR pulses in 50 $\mu$m length cavity. Dashed lines show the analytic model predictions for type-I and type-II regimes, points shows the results of numerical simulations, the symbols "+" and "x" distinguish SR pulses emitted at the two cavity facets. The carrier density is normalized at the transparency carrier density $n_t$.}\label {fig7_50um}
\end {figure}

 The type-I SR regime, in which the condensate of lower polaritons occupies the entire sample,  can be achieved in short cavities.  Fig.\ref{fig7_50um} shows predictions of analytic model for the SR pulse peak power in a cavity of only  50 $\mu$m long. The first and second critical densities for type-I and type-II  SR regimes in such short cavity are well separated from each other ($n^{(I)}_{cr}=11n_t$ and  $n^{(II)}_{cr}=14n_t$, respectively).
 Note that predictions of analytic model are in good agreement with the results of numerical simulations in the short cavities as well (Fig.\ref{fig7_50um}, points).

 \begin {figure}[tbp]
\includegraphics [width=7.5cm] {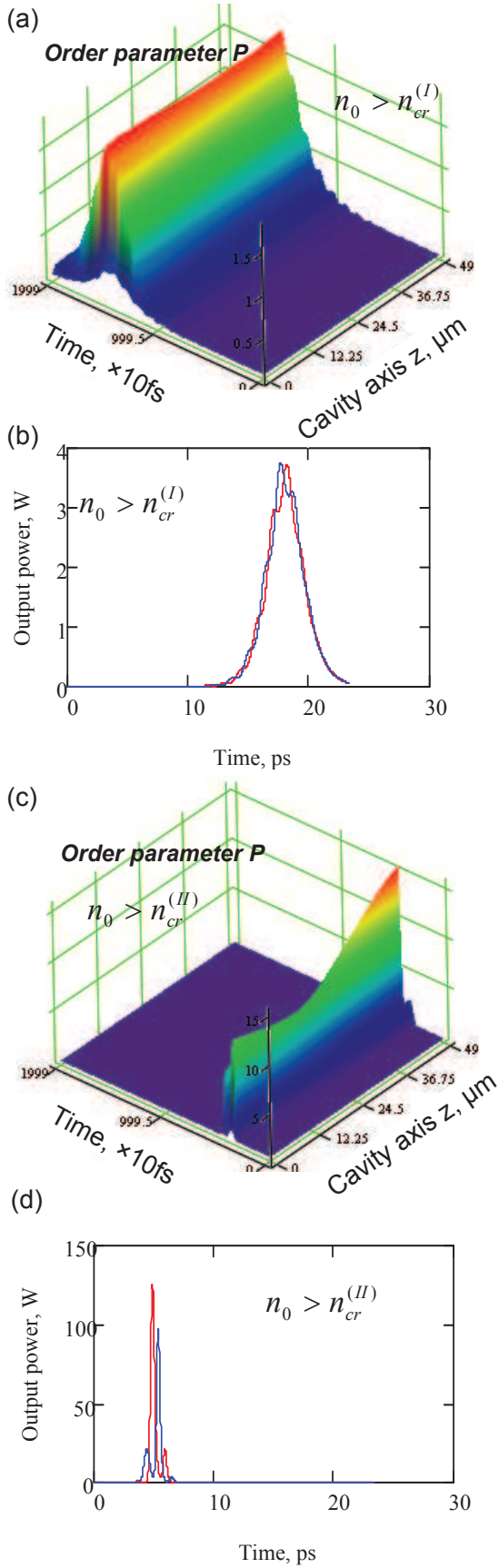} % use for Tex
\caption {Numerical simulations: Spatiotemporal dynamics of the order parameter $P$ [(a) and (c)] and SR pulses [(b) and (d)] in type-I [(a) and (b)] and type-II [(c) and (d)] regimes. The sample size is 50 $\mu$m. The initial carrier density $n_0$ is 13$n_t$ [(a) and (b)] and 28$n_t$  [(c) and (d)].
The SR pulse width is respectively  5 ps and 380fs, while the size of condensate fraction is 50 $\mu$m and 33 $\mu$m.}\label {fig8_50um}
\end {figure}

Figure \ref{fig8_50um} shows the results of numerical simulations for spatiotemporal dynamics of the order parameter and the output SR pulse shape in  the type-I and type-II regimes of condensation in this cavity.

For $n_0=13n_t$ (Figs. \ref{fig8_50um} (a) and (b)) the SR pulse width is 5 ps, much longer than the cavity roundtrip time (1.2ps). The SR pulse builds up after 15 cavity roundtrips. The order parameter $P$ reaches the maximum of 2$n_t$, in perfect agreement with the estimates of the analytic model of  $n_0-n^{(I)}_{cr}$ (see Eq.(\ref{IntMotion})). The SR pulse shape well agrees with hyperbolic secant shape assumed in the analytic model (Fig. \ref{fig8_50um} (b)). The coherence length is such that the condensate of LP occupies the entire sample of 50 $\mu$m long.

In the second example plotted in  Figs. \ref{fig8_50um} (c) and (d), the initial carrier density $n_0=28n_t$ well exceeds the critical density $n^{(II)}_{cr}$ and condensation of LP associated with emission of SR pulse undergos type-II transition. All stored energy is emitted in a single-shoot SR pulse after 3 cavity roundtrips,  when the order parameter $P$ reaches a maximum of 15$n_t$. The peak value of the order parameter during condensation of LPs is very close to predictions of the analytic model $n_0-n^{(II)}_{cr}=14n_t$. Small difference can be attributed to the ringing in the output SR pulse envelope, which  is not taken into account by analytic model.

The numerical simulations exhibit large amplitude and timing instabilities of the output SR pulses. The macroscopically large fluctuations are %one of
  the characteristic features of a condensate state \cite{Butov94,Altshuler91}. In this sense, the SR is a particularly interesting quantum optics phenomenon because the quantum fluctuations manifest themselves macroscopically, in the time and energy domains. %[15, 16].
 Of course, these features cannot be reproduced by analytic model based on Ginzburg-Landau master equation (\ref{GLE}).

The delay time statistics and timing jitter of SR pulses in atomic (molecular) ensembles have been thoroughly studied previously by both classical and quantum mechanical approaches.  In some conditions, the standard deviation of the SR delay time was found to be 10-12 \% \cite{Haake81,Watson83}. However, the results of previous studies can hardly be applied to our case of semiconductor laser structures. The reason for this is as follows.

One of the fundamental differences between SR in atomic (or molecular) gases and SR in semiconductors considered here consists in the opposite relation between the decoherence time $T_2$ of the medium and the characteristic width of superradiant pulse  $\tau_{c}$.
Indeed,  $ \tau_{c} \ll T_2$ in the classical case of SR emission in a gas medium. The SR pulse width $\tau_{c}$ is so large (up to few hundreds of nanoseconds) that the  coherent domain $L^*$ occupies entire sample. All  quantum oscillators contribute coherently to the field of the SR pulse,
 leading to
 the superposition $I \propto \left \langle \sum |E_i|^2 \right \rangle= n_0^2 \left \langle |E_i|^2 \right \rangle $. From this point of view, the SR emission in gases is similar to the type-I SR regime in semiconductors (Sec. \ref{SecSR1}).
However, in semiconductors,
the SR pulsewidth $\tau_{c}$ exceeds the natural decoherence time $T_2$ ($\sim$ 100 fs).  A more gentle growth of the output peak power with the initial carrier density in the  type-II SR regime (\ref{Tcoh2_ass}) has been ascribed to
 a partial coherence in the ensemble of e-h pairs during SR emission (Sec. \ref{SecTypeII}). Extending these considerations, we shall conclude that
 some part of the initially stored energy in the system as non-equilibrium electron-hole pairs of density $n_0$
  is spent to sustain the coherence of SR pulse against the dephasing processes.
Thus in Fig. \ref{fig345_TZDyn},
the SR pulses with ringing
evolve survive during a remarkably long time as compared to the
ultrafast intraband dephasing processes, while the invariant relationship (\ref{IntMotion}) is not perfectly fulfilled.
%it is remarkable how long fully developed pulses with ringing evolve over such long times despite of the presence of ultrafast intraband dephasing processes.
%%In Fig. \ref{fig345_TZDyn} the net modal gain after half roundtrip in the cavity is $g_0(n-n_t)\Gamma L / v_g\approx 40$.

The predictions of our model for the SR pulse width being longer than the inherent dephasing time of carriers in the InGaN/GaN QWs is in agreement with the previous
experimental studies of SR in GaAs devices \cite{Vasil'ev04,Vasil'ev99}, where the following considerations have been used \cite{MacGillivray76}.
%The SR emission from semiconductors exhibits the SR pulsewidths ($\sim\tau_{c}$) and coherency times to be much longer than $T_2$ determined by the intraband relaxation of individual e-h pairs ($\sim$100 fs).
%For the explanation of this contradiction,
MacGillivray and Feld discussed the effect of high optical gain on dephasing and relaxation times of the system and arrived to conclusion that
% According to Ref. [\onlinecite{Butov94}],
in the presence of the high optical gain the system can eventually dephase but with characteristic time  $T_2^{\rm{eff}}\sim T_2g_0(n_0-n_t)\Gamma L/ v_g$. In conditions of our numerical experiment, this effective dephasing time  is 4 ps.
%It is thus overcome dephasing during the early stages of SR pulse evolution. The ultrafast relaxation plays much stronger role in the performance of SR pulses (especially in terms of time and intensity fluctuations) from semiconductors as compared to this in gases and solid-state systems.
The macroscopic polarization of e-h system eventually dephases, but the effective dephasing time of the whole system is increased.

\section{Conclusion}

In summary, we present the analytical model of SR emission and condensation of 1D longitudinal polaritons in semiconductor laser cavities.
The master equation for the LP condensation and SR emission is shown to be analytically similar to the Ginzburg-Landau equation or Gross-Pitaevskii equation. The effective mass of condensing lower polaritons is very low, 10$^{-11}m_0$, giving a good reasoning for the thermal de Broglie wavelength criteria of  macroscopic quantum degeneracy do not appear in our GLE for SR.
We predicted  two regimes of SR emission and polariton condensation. We show that analytic model  predictions well agree with the results of the numerical simulations based on the traveling-wave Maxwell-Bloch equations. %

The effective mass of longitudinal 1D polaritons in  edge-emitting laser cavities is many order of magnitude smaller as compare to effective masses of 2D exciton-polaritons in semiconductor microcavity systems, rendering 1D edge-emitting laser structures particularly attractive for experimental observations of macroscopic coherent states at room temperatures.

 The work on incorporation of the finite interband relaxation time of carriers into the model is ongoing and will be published elsewhere. Other intersting development is related to the following.
 The GLE equation for BCS transition in electron system contains vector potential of the field, so as the critical condensation density is a function of the applied magnetic field. Similar effect in polariton system (\ref{GLE}) is under investigation now.

%\begin{acknowledgments}
\ack
This research is supported by the EC Seventh Framework Programme FP7/2007-2013 under the Grant Agreement n° 238556 (FEMTOBLUE)
%\end{acknowledgments}


\begin{thebibliography}{10}
\bibitem{CohExpl} In definition of the coherence time of the pulse, there is an element of ambiguity related to the intensity level at which the pulse width is measured. Thus
$2 \tau_p$ is the with of the pulse ${A_{+}^2(t{-}z{/}v_g)}$ at the level of $1/\rm{cosh}^2(1) {\approx} 42\% $, while the  FWHM is %$\tau_{FWHM}{=}$
$2\ln(1{+}\sqrt{2})\tau_p \approx 1.76 \tau_p$. The difference between two values is not fundamental.
\bibitem{SampleSize} Practical realization of SR emission from semiconductor laser cavity requires an intracavity absorber. In that case, the size of the sample treated by the theory is defined by the length of the gain section.
\end{thebibliography}
\end{document}